\documentclass[twocolumn]{aastex61}
\pdfoutput=1 
\usepackage{amsmath,amstext}
\usepackage[T1]{fontenc}
\usepackage{apjfonts} 
\usepackage[figure,figure*]{hypcap}

\usepackage{amssymb}
\input{epsf}
\usepackage{graphicx}

\usepackage{color}

\def\beq{\begin{equation}}
\def\eeq{\end{equation}}
\def\bea{\begin{eqnarray}}
\def\eea{\end{eqnarray}}

\shorttitle{The spin evolution of PSR J0537-6910}
\shortauthors{Andersson et al.}

\begin{document}

\title{The enigmatic spin evolution of PSR J0537-6910: r-modes, gravitational waves and the case for continued timing}
\author{N. Andersson}
\affiliation{Mathematical Sciences and STAG Research Centre, University of Southampton,
Southampton SO17 1BJ, United Kingdom}
\author{D. Antonopoulou}
\affiliation{Nicolaus Copernicus Astronomical Center, Polish Academy of Sciences, ul. Bartycka 18, 00-716 Warsaw, Poland}
\author{C.M. Espinoza}
\affiliation{Departamento de F\'isica, Universidad de Santiago de Chile, Estaci\'on Central, Santiago 9170124, Chile}
\author{B. Haskell}
\affiliation{Nicolaus Copernicus Astronomical Center, Polish Academy of Sciences, ul. Bartycka 18, 00-716 Warsaw, Poland}
\author{W.C.G. Ho}
\affiliation{ Mathematical Sciences \& Physics and Astronomy, STAG Research Centre, University of Southampton,
Southampton SO17 1BJ, United Kingdom}

\begin{abstract}
We discuss the unique spin evolution of the young X-ray pulsar PSR J0537-6910, a system in which the regular spin down is interrupted by glitches every few months. Drawing on the complete timing data from the Rossi X-ray Timing Explorer (RXTE, from 1999-2011), we argue that a trend in the inter-glitch behaviour
points to an effective braking index close to $n=7$, much larger than expected. This value is interesting because it would accord with the neutron star spinning down due to gravitational waves from an unstable r-mode. We discuss to what extent this, admittedly speculative, scenario may be consistent and if the associated gravitational-wave signal would be within reach of ground based detectors. Our estimates suggest that one may, indeed, be able to use future observations to test the idea. Further precision timing would help enhance the achievable sensitivity and we advocate a joint observing campaign between the   
Neutron Star Interior Composition ExploreR (NICER) and the LIGO-Virgo network.
 \end{abstract}

\keywords{stars: neutron
--- stars: rotation --- X-rays: stars --- gravitational waves --- pulsars: individual (PSR J0537-6910)}

\section{From  observation to  speculation} 

The young X-ray pulsar  PSR J0537-6910 in the Large Magellanic Cloud (associated with the supernova remnant N157B \citep{Xrdiscovery}) is an intriguing object. Spinning at a frequency of 62~Hz, this is the fastest spinning and most energetic non-recycled neutron star.  It also exhibits abrupt spin-ups (glitches) roughly every 100 days \citep{mid2006}.  Recent work has analysed this glitch activity \citep{0537_1,0537_2}, drawing on the complete timing data from the {\em Rossi X-ray Timing Explorer (RXTE)}. The results highlight the (almost) predictable regularity of the glitches, the overall (glitch dominated) spin-evolution and the inter-glitch behaviour. 

The analyses by  \citet{marshall,0537_1} and \citet{0537_2} also  raise the issue of the braking index of this neutron star. Assuming a spin-down rate that scales as $\dot{\nu}=-C\nu^n$, the braking index $n$ can be estimated as 
\beq
n=\frac{\nu\ddot{\nu}}{\dot{\nu}^2} 
\eeq
where $\nu$ is the spin frequency and dots denote time derivatives.
The data set points to a negative value of $n\approx -1.2$ over the observed 13 years \citep{0537_1}, a somewhat surprising result which may be related to the glitch activity. By focussing on the inter-glitch evolution one would infer much larger values for the braking index (see table 1 in \citet{0537_1}). It would be natural to assume that the observed phenomenology will ultimately be explained by the detailed nature of the glitch relaxation, involving superfluid vortex dynamics and poorly understood friction/pinning forces. However, there may be additional physics at play. 

As we will discuss, the long-term post-glitch relaxation hints at the system evolving towards an effective braking index of $n\approx7$. Such a large value would suggest that the spindown of J0537-6910 is not governed by electromagnetic emission. Taking the result at face value we are instead led to consider the possibility that the star spins down due to gravitational radiation. However, even in this case we have to go beyond the ``standard'' scenario. A spinning deformed star would emit quadrupole radiation at twice the spin frequency, leading to a braking index $n=5$. Not what we are looking for. There is, however, a plausible scenario that would ``explain'' the timing data. If the gravitational-wave driven instability of the inertial r-mode \citep{nareview,revolve} were to operate in the star, and the associated emission were to dominate the spin evolution, then theory predicts (for the quadrupole mode, which mainly radiates through the mass current multipoles) a braking index of exactly $n=7$. While this may be a coincidence, and the true explanation for the spin evolution of J0537-6910  lies elsewhere, the possibility is interesting enough that it warrants a more detailed discussion.

In this paper we take a closer look at the RXTE timing data, with the aim of establishing to what extent the suggested braking index of $n\approx 7$ is credible and robust. Not surprisingly, the answer will be inconclusive. Next we consider whether there is a workable r-mode scenario for this system. Again, we can not draw definite conclusions, but our discussion highlights the parts of the theory one might have to negotiate in order to arrive at a consistent model. This naturally leads us to the issue of future observations. We provide simple estimates that suggest that the gravitational waves associated with the r-mode scenario should be within reach of the advanced generation of interferometers. In essence, one may be able to use observations to constrain (and perhaps rule out) the presence of an unstable r-mode in this system. 

In order to achieve this, one would ideally carry out a targeted search for r-mode gravitational waves from J0537-6910. This requires reliable timing information.
This is crucial as the frequent glitches disrupt predictions made by timing models developed prior to each glitch.  However, the last RXTE  timing observations of J0537-6910 took place in 2011. This means that the system was not considered in targeted LIGO searches in the advanced detector era (see \citet{ligos6} for the most sensitive search, at twice the spin frequency, and \citet{ligoO1} for the best current results for other pulsars). Given that  simultaneous gravitational-wave and X-ray observations would
enable the most effective gravitational-wave search, we argue that there is a strong case for the advanced LIGO-Virgo network to join forces with   the {\it Neutron Star Interior Composition ExploreR} ({ NICER}), an X-ray telescope recently installed on the International Space Station \citep{gend}.
{\it NICER} is optimized for detecting pulsations from neutron stars such as
J0537$-$6910, having twice the collecting area of {XMM-Newton} and a timing
accuracy of 100~ns.  With a careful observing strategy,
{NICER} should be able to track the spin evolution of J0537-6910, including detecting
glitches, thus allowing gravitational-wave searches to compensate for the complex
timing behaviour.

\section{The observed rotational evolution}

\begin{figure}[h]
\begin{center}
\includegraphics[width=0.9\columnwidth,clip]{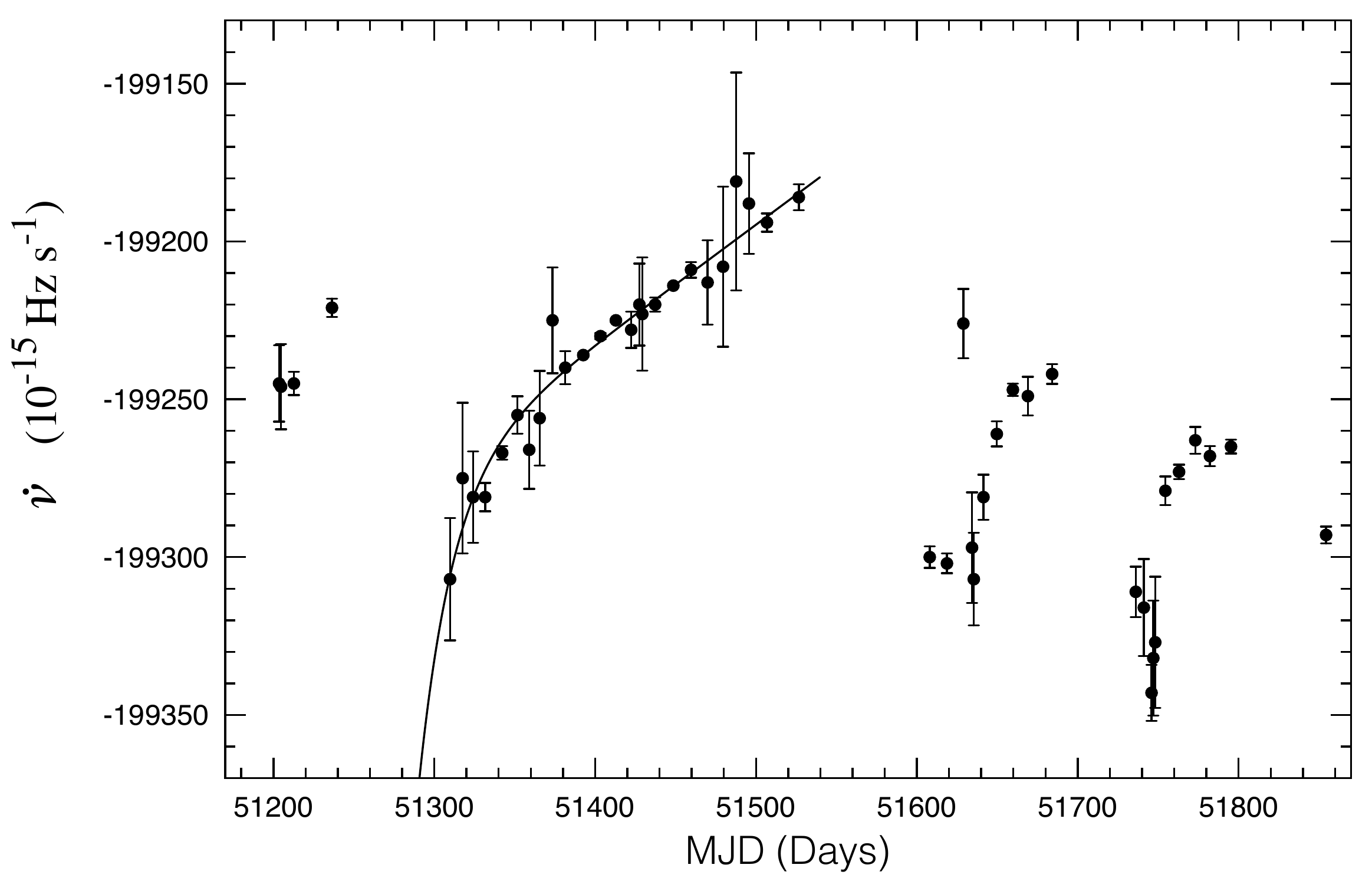}
\end{center}
\caption{An illustration of the evolution of $\dot{\nu}$ through the first two years of observations, including 4 glitches (points are calculated as in Fig.5 of \citet{0537_1}). 
The solid line shows a model of the form 
$\dot{\nu}_0 + \ddot{\nu}_0(t-t_g) - (\Delta\nu_d/\tau_d) e^{-(t-t_g)/\tau_d}$, with the best-fitted parameters derived from a fit to all ToAs following the first glitch, which occurred near $t_g= {\rm MJD}\, 51278$, and up to the next glitch (see text for details). 
The linear term of this model, $\ddot{\nu}_0 =(4.4\pm0.1)\times10^{-21} \; \rm{Hz\,s^{-2}}$, implies an underlying braking index of $6.8\pm0.2$. } 
\label{expo}
\end{figure}

PSR~J0537--6910 has a unique rotational evolution: it abruptly spins up by a few ppm every few months,  by far the highest glitch rate observed in any pulsar \citep{fuentes}. The imprint of the frequent glitches in the spin-down rate is  dramatic. Most spin-ups $\Delta\nu$ are accompanied by a sharp decrease in $\dot{\nu}$ and a subsequent recovery characterised by a large, positive $\ddot{\nu}$ (see figure~\ref{expo} for a sample of the data). The slope of this saw-like pattern over the course of $13$ years returns a long-term negative $\ddot{\nu}$ and an apparent braking index of $n=-1.22\pm0.04$ \citep{0537_1}.  

It is far from trivial to differentiate the effect of glitches from the underlying braking mechanism. 
This is mainly because the time intervals between glitches are short (only a few months long), while the timescales associated with the superfluid response to glitches might be of the order of years \citep{bryn14}.  Therefore, we can not guarantee  that the internal superfluid has  reached its equilibrium state at the time of a given  $\ddot{\nu}$ measurement.  

Between glitches, it is possible to carry out a phase-coherent timing analysis to derive the rotational parameters. 
The rotational phase of the pulsar, $\phi(t)$, can be described by a truncated Taylor series around an epoch $t_0$
\beq
\label{BasicTimingModel}
\phi(t)=\phi_0+\nu_0(t-t_0)+\frac{\dot{\nu}_0}{2}(t-t_0)^2+\frac{\ddot{\nu}_0}{6}(t-t_0)^3
\eeq 
where $\phi_0$, $\nu_0$, $\dot{\nu}_0$ and $\ddot{\nu}_0$ are the reference phase, spin frequency and its first two time derivatives, respectively. 
When this simple timing model is fitted to  entire inter-glitch intervals, the inferred braking indices (calculated as $n=\nu_0\ddot{\nu_0}/\dot{\nu}_0^2$)  are typically greater than $10$. 
Such large values of $n$ most likely reflect the early response to the glitch. 
The analysis of \citet{0537_1} shows that $\ddot{\nu}_0$ (or equivalently, $n$) tends to be smaller for longer interglitch intervals, as data further away from a glitch start to dominate the fits. 

In fact, following the first (and largest observed) glitch, a simple analysis gives an ``average'' braking index of $n=7.6\pm0.1$ (see table~1 in \citet{0537_1}). 
This particular inter-glitch time interval is much longer than any other ($\sim284$ days, while all others are less than $200$ days) and an exponential relaxation on a relatively short, 20-day timescale was observed. 
The estimate of $n$ in \citet{0537_1} was derived by fitting only equation \eqref{BasicTimingModel} to data after the first 16 days post-glitch (to avoid the quickly decaying initial phase) and up to the second glitch. 
To examine the asymptotic $\ddot{\nu}$ of this interglitch interval in more detail, we need to account for the exponentially decaying term in the timing solution (see figure~\ref{expo}). 
We thus fit all available pulse times-of-arrivals (ToAs) between the first and second glitch with a timing model, as in equation \eqref{BasicTimingModel}, including an exponential term. 
The parameters of the additional term are consistent with those in Table 3 of \citet{0537_1} and we find the underlying braking index to be $n=6.8\pm0.2$.
 
Given the possible connection between this value for the braking index and a scenario involving gravitational-wave emission through unstable r-modes, we want to further investigate the inter-glitch braking indices and their apparent softening at late times. Basically, we want to check whether a braking index close to 7 is unique to the evolution after the first glitch, or could be an asymptotic behaviour common to more glitches.  
The details on the observations, derivation of the ToAs and pulsar timing tools used in the following can be found in \citet{kuiper} and \citet{0537_1}. 

To calculate the braking index at different moments after each glitch, we fitted equation \eqref{BasicTimingModel} to short segments of data. 
Because $\ddot{\nu}_0$ is small, of the order of $10^{-20}\;\rm{Hz\,s^{-2}}$, its measurement can be largely affected by noise (ToA uncertainties, but also ``timing noise'' intrinsic to the pulsar). This compromises the accuracy of the best-fitted values, especially for time intervals shorter than about $50$ days or ones that contain too few ToAs. 
For the shortest inter-glitch intervals, the entire data had to be used and only one measurement of the braking index was obtained. 
For the longer intervals, however, it was possible to get a few measurements at different epochs. 
We used segments of interglitch data spanning just below 90 days, with the precise length depending on the distribution of the ToAs, and imposed a minimum of at least 5 ToAs for each fit. 
Several values of the interglitch $n$ for different times after the glitch were calculated by shifting (when possible) the fitting window by 20 days.

\begin{figure}[h]
\begin{center}
\includegraphics[width=0.9\columnwidth,clip]{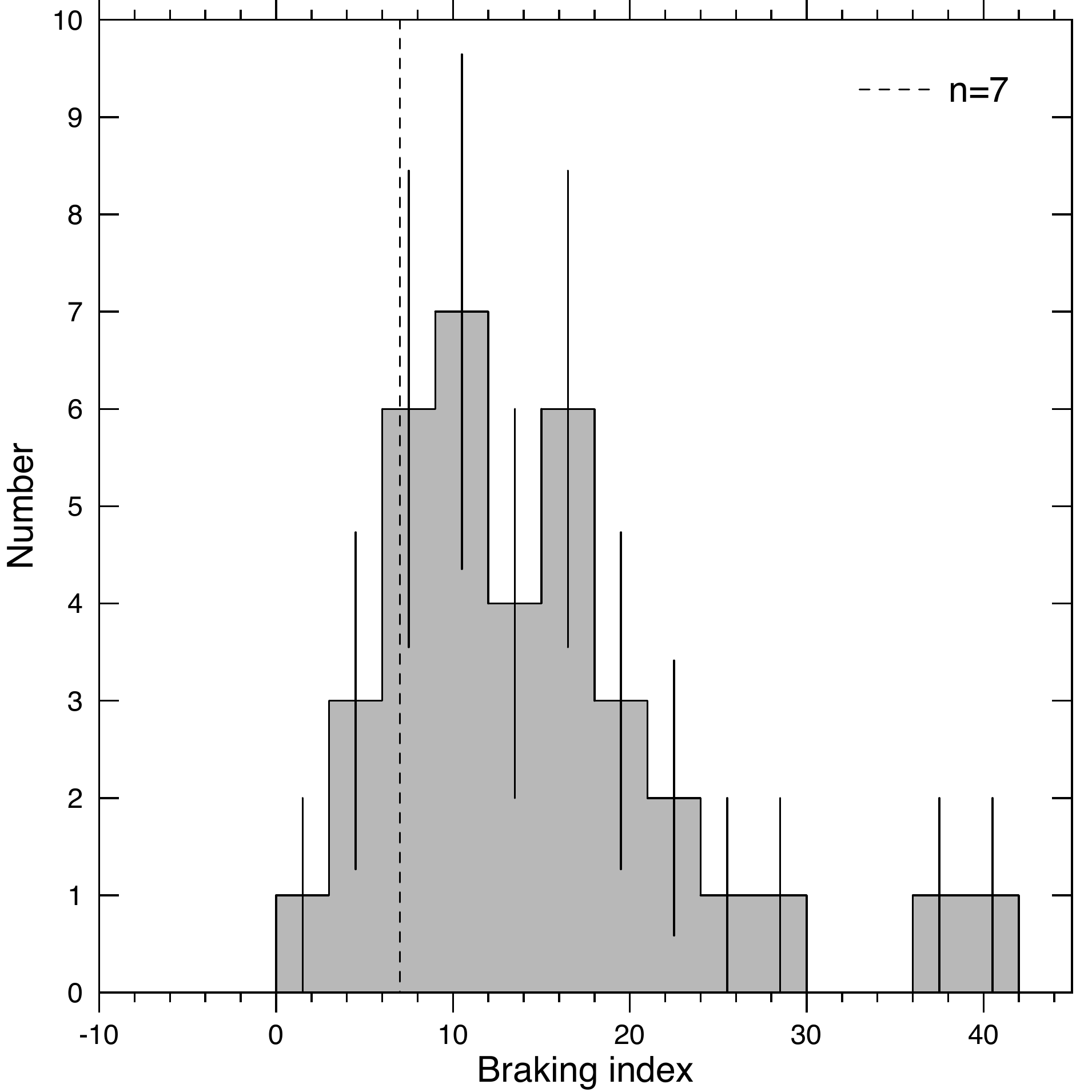}
\end{center}
\caption{The braking index obtained from direct fits of $\nu_0$, $\dot{\nu}_0$ and $\ddot{\nu}_0$ to inter-glitch ToAs. Only values from fits centred at least 50 days after each glitch are shown, which excludes some of the shortest inter-glitch intervals. For the longest inter-glitch intervals, segments of data $\leq 90$ days long were used for the fit, centred approximately $20$ days apart (see text for details, and figure \ref{n_time} for the evolution of $n$ as a function of  time since the glitch).   } 
\label{histo}
\end{figure}

\begin{figure*}
  \includegraphics[width=\textwidth]{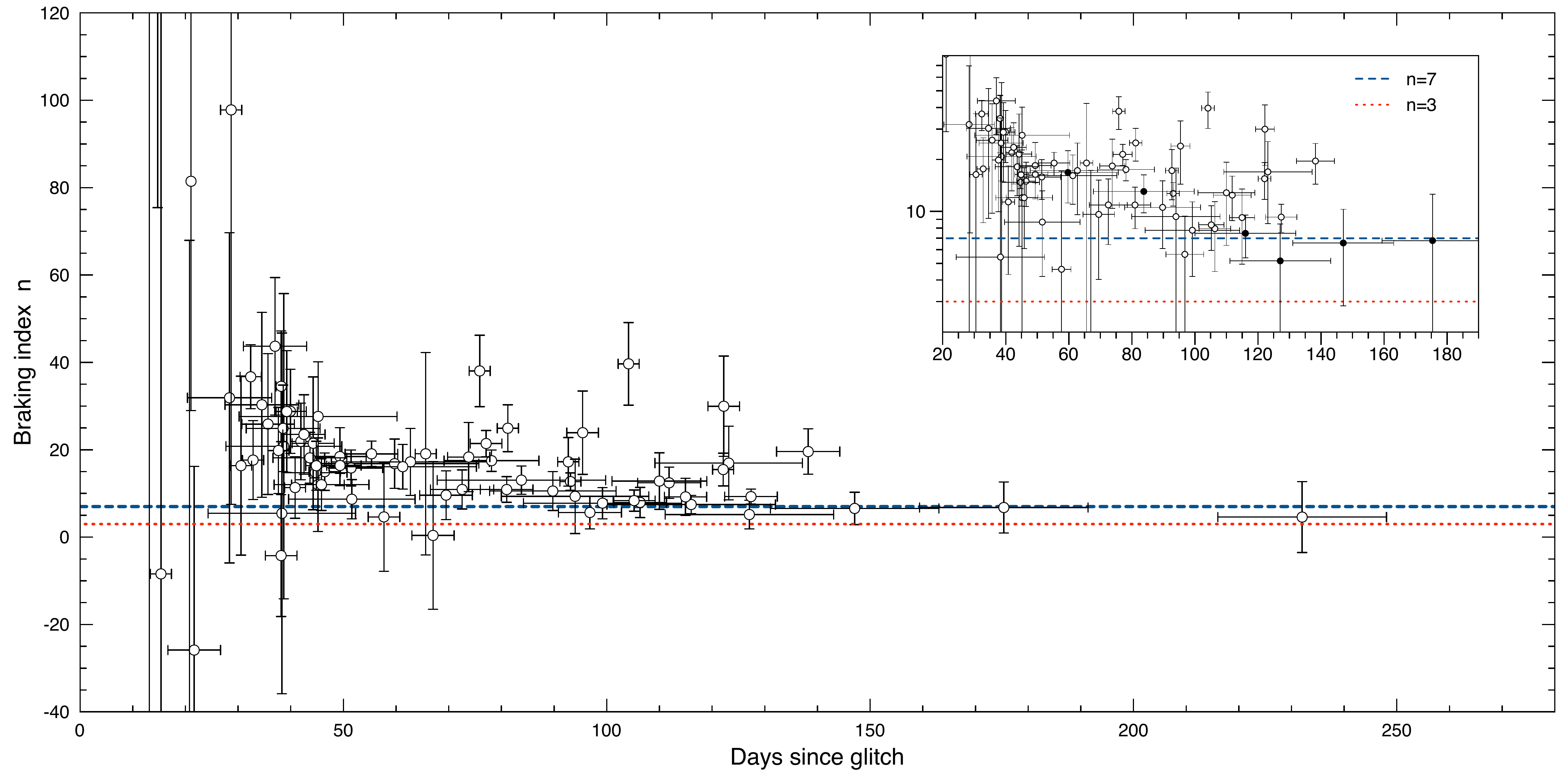}
  \caption{The braking index $n$ as a function of time $t_{\rm pg}$ since the preceding glitch. The data points are obtained from direct fits of \eqref{BasicTimingModel} over a sliding window up to $90$ days long, which was moved forward by 20 days at each step. The dashed horizontal lines indicate values of $n=3$ (as expected for a spin down dominated by electromagnetic dipole radiation) and  $n=7$ (which would apply in the scenario explored in this paper). The insert presents a zoomed-in plot of the same data in logarithmic scale, with the data points corresponding to fits of the first interglitch interval highlighted in black.}
  \label{n_time}
\end{figure*}

A histogram of the results for the braking index is shown in Figure~\ref{histo}. For clarity, we have excluded fits centred on ToAs that are less than 50 days away from the glitch epoch, as these are dominated by the early fast relaxation. Typically, larger values of $n$ correspond to epochs soon after a glitch as demonstrated in Figure \ref{n_time}, which shows the braking index $n$ as a function of time since the preceding glitch, $t_{\rm pg}$. This kind of plot, however, has to be considered with some caution. As reference dates, we have used the MJD epoch for the 45 glitches reported in \citet{0537_1}. The errors from the uncertainty in the glitch epoch are displayed in Figure 2. Using a slightly different set of ToAs (derived from the same {\it RXTE} data) \citet{0537_2} obtain very similar results for the glitch parameters, but in at least 3 cases the glitch epochs of the two datasets are inconsistent within the errorbars -- errors in figure 2 should thus be viewed as lower limits. Errors in the braking indices are propagated 1-sigma uncertainties on the best-fitted parameters of \eqref{BasicTimingModel}, which are often underestimates. Moreover, although we believe the list of 45 glitches to be complete for spin-ups larger than $\sim1 \,\rm{\mu Hz}$, some small events may not have been identified. For example, \cite{0537_2} discovered another possible glitch at MJD 52716(1).  Missing glitches may  introduce an overestimate of $t_{\rm pg}$. The opposite would be true if some of the timing irregularities considered as glitches were in reality timing noise features\footnote{A total of 4 events were flagged as ambiguous by \citet{0537_1}. We recalculated $t_{\rm pg}$ under the assumption that these timing features were not real glitches and confirmed that their inclusion does not alter the main features of Figure \ref{n_time}.}. These are important caveats, but the qualitative picture of the braking index evolution remains the same. There is some tendency towards a value of $n$ close to 7 as $t_{\rm pg}$ progresses, although a robust measurement is possible only for the first glitch (and longest glitch-free interval).

A similar conclusion was reached by \citet{0537_2} following a rather different approach. They identified 42 glitches and used a subset of them to construct $\dot{\nu}$  as a function of $t_{\rm pg}$. They then combined those values to achieve a set of denser data points which they fit with a single function $\dot{\nu}(t)$, assuming that the relaxing component of all glitches is described by a single exponential with the same amplitude and characteristic timescale. Using the best-fitted $\ddot{\nu}$ from \citet{0537_2}, we infer that the asymptotic value of the braking index for the inter-glitch time intervals is $n=7.4\pm0.7$. This is consistent with the results shown in Figure~\ref{n_time}.

These arguments suggest that an underlying $n\approx 7$, inferred for the evolution after the first glitch, might be potentially accommodated by the entire 13 years of data. 
Whether this is close to the ``real'' long-term braking index, and thus probes the dominant braking mechanism is open to interpretation. In the standard glitch scenario, the recovery is governed by the microphysics of the internal superfluid and can often extend for a very long time. It would thus not be surprising if most glitches were to exhibit a similar relaxation, consistent with a braking index around $7$ after  some time post-glitch, but which would gradually decrease further if the next glitch did not interrupt the process. 

This issue could possibly be resolved by future observations of one or more glitches larger than the first event in the RXTE data (or at least, of similar size). Since the size of each J0537-6910 glitch strongly correlates with the time interval to the next one, larger glitches would enable us to populate the region beyond $t_{\rm pg}>150$ days with more $n$ measurements. This would allow us to assess the significance of the observed tendency. 
As such big spin-ups appear to be at the higher end of the glitch size distribution for this pulsar, they are likely relatively rare events. It is therefore important that if such a glitch is observed, the post-glitch relaxation is monitored closely. Frequent observations are not only needed for accurate $n$ calculations, but will moreover provide a way to test the hypothesis in \citet{0537_2} that the parameters of the exponential recovery are common to all glitches. 

\section{Evidence of an unstable r-mode?}

Turning to the possible explanation for the observed behaviour, let us first go through the argument that leads to the braking index for a star that spins down due to an unstable r-mode.  In general, the r-mode instability arises through a tug-of-war between gravitational-wave emission, which drives the instability, and various dissipation mechanisms, that saps energy from the mode. 
Somewhat schematically (adopting the strategy from \citet{revolve}) the amplitude of the mode $\alpha$ evolves according to 
\beq
\dot \alpha =  \alpha \left( {1 \over t_\mathrm{gw} }- {1\over t_\mathrm{diss}} \right) - {N\over 2I\Omega}
\label{alpeq}\eeq
The evolution depends on the instability growth time $t_\mathrm{gw}$ and the dissipation timescale $t_\mathrm{diss}$ (we take both timescales to be positive, in contrast with \citet{revolve}), as well as any external torque $N$. Letting  $I$ be   the star's moment of inertia
and $\Omega$ the angular spin frequency (we assume uniform rotation for simplicity), we also have
\beq
\dot \Omega = - {2Q \Omega \alpha^2 \over t_\mathrm{diss}} + {N\over I}
\label{omeq}\eeq
where $Q$ is an equation of state dependent quantity.
In this phenomenological model,  the unstable r-mode is assumed to grow exponentially until it reaches a given saturation amplitude $\alpha_s$. From equation~\eqref{alpeq} we see that, if $\alpha_s$ takes a fixed value then (assuming that we can ignore external torques) we must have $t_\mathrm{diss}\approx t_\mathrm{gw}$. It then follows from equation~\eqref{omeq} that the spin frequency $\nu=\Omega/2\pi$ evolves according to
\beq
\dot  \nu = - {2Q\alpha_s^2 \nu \over t_\mathrm{gw}} 
\label{nueq}
\eeq
If we, in order to keep things simple enough that the scalings with the star's mass $M$ and radius $R$ are apparent, consider an $n=1$ polytrope, then the growth time for the $l=m=2$ r-mode is given by (all r-mode timescale estimates are taken from the review by \citet{nareview})
\beq
t_{\rm gw} \approx  5\times10^{7}  \left( {M \over 1.4M_\odot}\right) ^{-1} \left( {R \over 10~\mathrm{km}}\right)^{-4} \left( {\nu\over 100~\mathrm{Hz}}\right)^{-6} \ \mbox{ s}
\label{gwest}\eeq
Noting that we have  $Q \approx 9.4\times10^{-2}$  \citep{revolve}, it follows that 
\beq
\dot  \nu  \approx - 4\times10^{-7} \alpha_s^2  \left( {M \over 1.4M_\odot}\right) \left( {R \over 10~\mathrm{km}}\right)^{4}  \left( {\nu \over 100\ \mathrm{Hz}}\right)^7 \ \mathrm{s}^{-2}
\label{nueq}
\eeq
and we  arrive at the braking index $n = 7$,
as long as $\alpha_s$ is constant. 

The theory prediction thus accords with the behaviour inferred from the inter-glitch evolution of J0537-6910. This is an interesting observation, but it does not mean that we are done. We need to check to what extent this explanation is consistent. This involves considering poorly known aspects of neutron star interior physics but we should nevertheless be able to make some progress. At each step, we need to be mindful of the assumptions. So far, we have i) ignored external torques, which means that the gravitational-wave emission dominates the spindown, ii) assumed that the unstable r-mode has reached saturation, and that this corresponds to a constant amplitude $\alpha_s$. 

Given these two assumptions, we can combine \eqref{nueq} with the observed spin parameters for the system. 
Once we consider a specific neutron star model for which we can (at least in principle) work out the instability growth time, we can turn the data into a statement about the required  saturation amplitude. 

First of all, the estimated growth time \eqref{gwest} immediately tells us that, 
for a ``canonical'' neutron star with $R=10$~km and $M=1.4M_\odot$ spinning at the observed
$\nu=62$~Hz, the growth time would be $t_{\rm gw} \approx 30$~years. That is, the mode would not quickly regrow if it were disrupted by the frequent glitches. We also see that we need to keep an eye on the relatively high power of the stellar radius. If we consider the current radius constraint from X-ray observations, $R=10-14$~km \citep{stein}, then the growth time would be shorter by about a factor of 4 for the largest neutron stars. 

Next, making use of the observed $\dot \nu = 1.99\times10^{-10}$~Hz/s, we see that an r-mode dominated spin down requires 
\beq
\alpha_s \approx 0.12   \left( {M \over 1.4M_\odot}\right)^{-1/2} \left( {R \over 10~\mathrm{km}}\right)^{-2} 
\eeq
That is, if we consider the radius to be 14~km, then the required amplitude is about a factor of 2 smaller. 

If we assume that the star has been spinning down according to \eqref{nueq} throughout most of its history, and that it was initially spinning much faster than it is today, we can estimate the age. This way, we find that the evolution to the current spin-rate, $\nu$,  takes place on a timescale
\beq
t_\mathrm{sd} \approx 4.2\times10^{9} \left( {\alpha_s \over 0.1}\right)^{-2} \left( {\nu \over 100\ \mathrm{Hz}}\right)^{-6}\  \mathrm{s}
\label{spintime}
\eeq
If we let $\alpha_s$ have the predicted value then  it would take 1,600-6,000 years for a star with radius in the range $10-14$~km to reach the current spin rate. This is in good agreement with the estimated age of the supernova remnant, 1-5,000 years \citep{wanggotthelf98,chenetal06}.

However, given what we think we know about the mechanism that determines the r-mode saturation, the estimated $\alpha_s$ is uncomfortably large. It is generally expected that the nonlinear coupling between the large scale r-mode and the sea of shorter wavelength inertial modes will lead to saturation at $\alpha_s < 10^{-2}$ \citep{arras}. Is this a fatal objection to the proposed scenario? Possibly, but one can imagine ways of reducing the tension. First, it could be that the growth timescale is shorter than the estimate in \eqref{gwest}. We know, for example, that the result for a uniform density star is about a factor of 2 smaller than \eqref{gwest} \citep{nareview}. But this is not enough to make the results consistent. Instead we may consider the saturation mechanism. The level of saturation is expected to be close to the threshold where the nonlinear coupling between the r-mode and a pair of inertial daughter modes becomes parametrically unstable \citep{arras}. The threshold amplitude depends on the damping rates of the (supposedly stable) daughter modes and the level of frequency detuning (how close the mode frequencies are to resonance). Focussing on the former of these factors, we note that an increase in the daughter mode damping rate by some factor would affect the r-mode saturation by the same factor. That is, a more efficient damping of short wavelength daughter modes could bring the theoretical estimate closer to the required saturation amplitude. The question is if this is reasonable.

In order to discuss this issue, we need to consider the possible dissipation mechanisms that may act on an unstable r-mode. In the simplest model, the dominant dissipation is the macroscopic shear viscosity due to neutron-neutron scattering. In this case we have a damping timescale \citep{nareview}
\beq
t_{\rm sv} \approx 6.7\times10^5 \left( {M \over 1.4M_\odot}\right)^{-5/4} \left( {R \over 10~\mathrm{km}}\right)^{23/4} \left( {T \over 10^8~\mathrm{K}}\right)^2\ 
  \mbox{ s} 
\label{sv1}
\eeq

In order for the star to be in the unstable regime, so that the previous arguments hold, we need the damping rate to be slower than the growth rate from \eqref{gwest}. This leads to the condition;
\beq
T > 8.6\times10^8 \left( {M \over 1.4 M_\odot}\right)^{1/8} \left( {R \over 10~\mathrm{km}}\right)^{-39/8} \left( {\nu \over 100\ \mathrm{Hz}}\right)^{-3}\ \mathrm{K}
\label{thres}
\eeq
The weak scaling with the star's mass means that the dependence on the radius is  dominant.
For the observed spin-rate $\nu=62$~Hz, we would need $T>3.6\times10^9$~K for a 10~km star. This is uncomfortably hot. However, if we take the radius to be 14~km, then we only need  need $T>7\times10^8$~K. As a useful comparison, we note that one would expect the star (in absence of any instability) to have cooled to a temperature of roughly $T\approx 2\times 10^8$~K \citep{sciad}.
The relatively high temperature threshold simply reflects the expectation that we may need a larger than anticipated r-mode instability window in order to accommodate this system.

\begin{figure}[h]
\begin{center}
\includegraphics[width=0.9\columnwidth,clip]{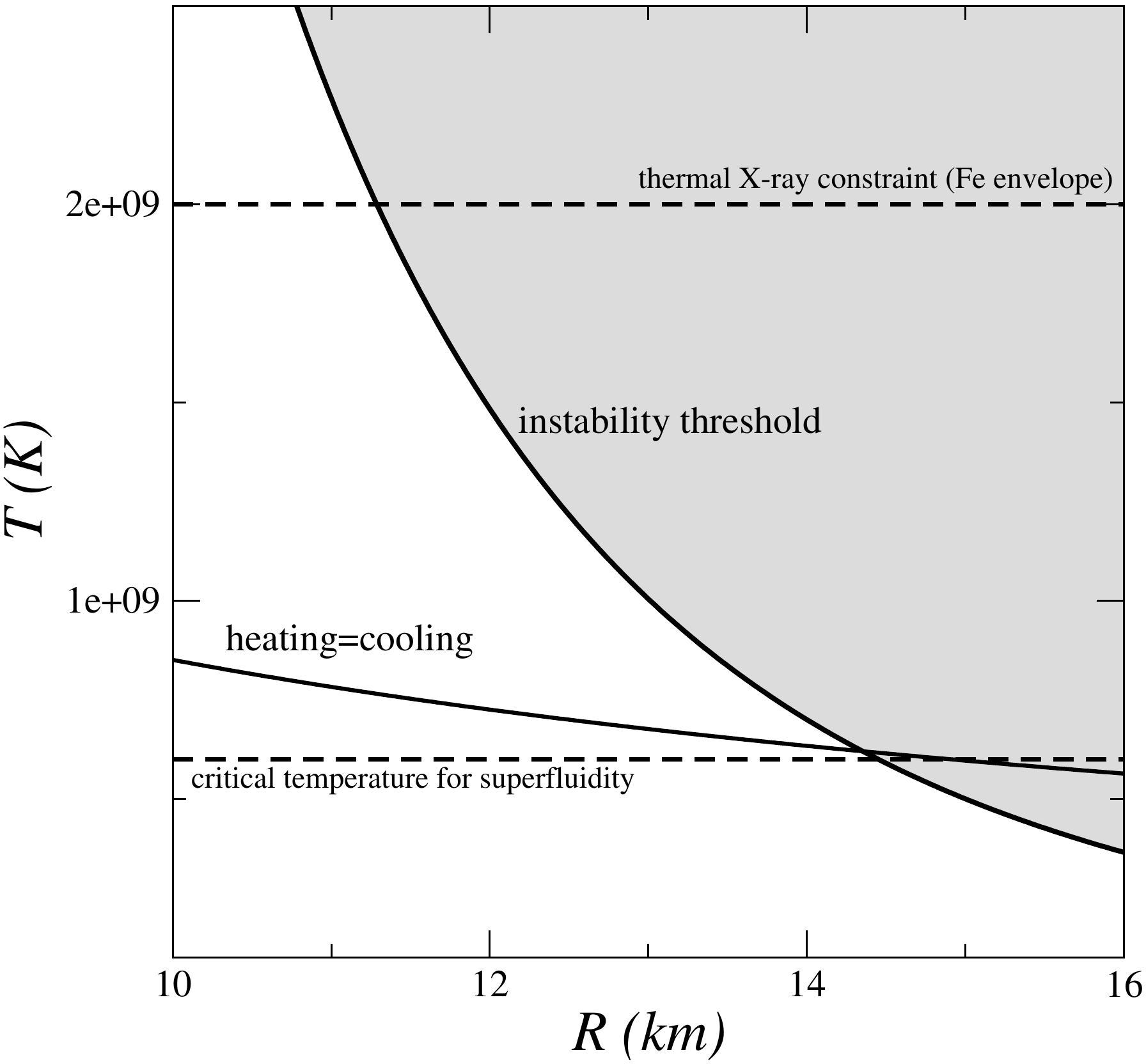}
\end{center}
\caption{Comparing the two temperatures from \eqref{thres} and \eqref{Tbal}. The first indicates the threshold above which the r-mode would be unstable. The second provides the temperature at which r-mode heating balances modified Urca cooling. The results suggest that we need the neutron star radius to be greater than about $14.5$~km in order for the system to be at thermal balance inside the r-mode instability window. However, one would not have to change our estimated growth/damping timescales by much to bring the radius into the suggested range of 10-14~km. We also show the X-ray constraint on the star's core temperature (which sets an upper limit), assuming a heavy element (Fe) envelope, as well as an indicative level for the onset of core superfluidity (with many viable models entering at a lower level than this).  } 
\label{temps}
\end{figure}

However, the model has an internal ``consistency check''. The friction associated with the viscosity heats the star. After some time of evolution, one would expect this heating to balance to cooling due to neutrino emission. This balance dictates the star's thermal evolution. The mode heating follows from
\beq
\dot E_\mathrm{sv} = {\alpha^2 \tilde J \Omega^2 MR^2  \over t_\mathrm{sv}}
\eeq
For an $n=1$ polytrope, we have $\tilde J \approx 1.635\times10^{-2}$ \citep{revolve} so
\begin{multline}
\dot E_\mathrm{sv} \approx 2.7\times10^{43} \alpha^2_s \left( {M\over 1.4 M_\odot}\right)^{9/4}  \left( {R\over 10\ \mathrm{km}}\right)^{-15/4} \\
\times \left( {\nu\over 100~\mathrm{Hz}}\right)^{2}\left( {T \over 10^8\ \mathrm{K}}\right)^{-2}\ \mathrm{erg/s}
\end{multline}
This should be compared to the neutrino luminosity associated with the modified Urca reaction. Using the estimated neutrino luminosity from \citet{shap} we have (for a constant density star) 
\beq
\dot E_\mathrm{mU} \approx 5.6\times10^{31} \left( {M\over 1.4 M_\odot}\right)^{2/3}  \left( {R\over 10\ \mathrm{km}}\right) \left( {T \over 10^8\ \mathrm{K}}\right)^8\ \mathrm{erg/s}
\eeq
Equating the two rates, for the observed spin rate, we see that thermal balance implies a core temperature;
\beq
T\approx 1.3\times10^9 \alpha_s^{1/5} \left( {M\over 1.4 M_\odot}\right)^{19/120}  \left( {R\over 10\ \mathrm{km}}\right)^{-19/40}\ \mathrm{K}
\label{Tbal}
\eeq
Making use of the inferred saturation amplitude from the spin down, we see that 
for a 10~km star, thermal balance implies that $T\approx 8.6\times10^8$~K. If we take the radius to be 14~km, the corresponding temperature is $T\approx 6.4\times 10^8$~K. 
In order to see if the model is consistent, we compare the two temperatures from \eqref{thres} and \eqref{Tbal} in Figure~\ref{temps}. The figure shows that we would need the radius of the star to be greater than about $14.5$~km in order for the system to be at thermal balance inside the r-mode instability window. Of course, one would not have to change our estimated growth/damping timescales by much to bring the radius into the suggested radius range (below 14~km). Basically, it may not be unreasonable to suggest that the instability could operate in this system.

The various estimates we have used obviously come with a range of caveats and one should perhaps not read too much into the conclusion that the r-mode scenario would appear to be consistent. We have based the argument on simple Newtonian estimates, which may be adequate for a first attempt but which cannot be used in combination with a more realistic equation of state. If we want to make the model more realistic, then we  have to consider the r-mode problem in general relativity \citep{lock1,lock2}. As an alternative, one could make use of parameterised versions of the different timescales, as advocated by \citet{alford}. 

Based on our estimates, the most important issue relates to the saturation amplitude, $\alpha_s$, which is at least an order of magnitude larger than predicted. Next, it is legitimate to ask if it is appropriate to rely on the neutron shear viscosity. One important issue regards the expected onset of  neutron superfluidity, which would suppress the shear viscosity. Below the superfluid transition temperature, the main shear viscosity is due to electron-electron scattering. If we had used the corresponding damping timescale (see \citet{nareview}) in our estimates then the different temperatures would not be consistent. However, it could well be that the outer core of the star has yet to cool below the superfluid transition. From the sample of relevant pairing gaps considered in \citet{casapairing} (see their figure~10) we learn that only the models with the largest gaps have a critical temperature above $6\times10^8$~K.  As our estimated temperatures are (just) above this, it does not seem unreasonable to assume that the neutrons are normal and hence that \eqref{sv1} applies. The fact that the composition of the neutron star core is uncertain is also a concern, but the main r-mode damping is associated with the fluid motion at around 70-80\% or so of the star's radius. As more exotic phases (and states) of matter may not be present in the star's outer core, it seems entirely plausible that their presence (or absence) at higher densities would have little effect on the r-mode damping. 

Before we proceed, let us make one further comment on the temperature. The core temperatures we require for the r-mode instability to be active are higher than one would expect if the star was simply cooling in isolation. Hence, it is worth considering whether one may be able to use X-ray observations to constrain the  scenario. Taking the observed non-thermal X-ray luminosity of $6\times10^{35}$~erg/s from \citet{chenetal06} as an upper limit on the surface emission, one would infer a surface temperature of about $5\times10^6$~K.  This converts into a limit on the core
temperature of $< 2\times10^9$~K (Fe envelope) or $<6\times10^8$~K (for H). The core temperatures we have inferred would be easily compatible with a heavy element envelope.

Finally, as an alternative, one may consider the possibility that the r-mode is stable (as one might have expected in the first place), but that it is excited by some impulsive mechanism. However, it is not straightforward to make a stable r-mode scenario consistent with the observations.  In order to arrive at the suggested braking index of $n=7$ we need the gravitational-wave emission to dominate the spin down and the result requires a constant mode amplitude. While one can easily evolve the mode amplitude in the stable regime, the result tends to be very different from what we require. Moreover, the impulsive mode excitation is problematic. One would need to pump a lot of energy into the mode, much more than seems allowed by the energy budget associated with the glitches. The unstable r-mode scenario resolves these issues in a seemingly natural way.

While our arguments suggest that the unstable r-mode scenario may accommodate J0537-6910, it is important to keep in mind that this would involve a large instability region at the inferred (relatively high) temperatures. The physics may, in principle, allow for this, but there is an obvious tension between this scenario and the many observed fast spinning accreting neutron stars in low-mass X-ray binaries \citep{andlmxb,hask2012,stroh,patruno}. 
In order for those systems not to spin down due the r-mode instability, the threshold must be around 600~Hz at a core temperature  a factor of a few lower than \eqref{Tbal}. This would require the instability to have a sharp feature in a fairly narrow temperature range. This could be problematic, but one can think of scenarios that would predict this behaviour. For example, the onset of core superfluidity, which brings vortex-mediated mutual friction into play, may have exactly this effect (see for example
 figure 6 in \citet{brynsf}).

\section{Observational tests}

The observational evidence and the theoretical estimates clearly do not settle the issue, but we can clearly not rule out the notion that the r-mode instability may impact on the spin-evolution\footnote{Note that a comprehensive model must also be able to accommodate the long-term behaviour of the pulsar, which is governed by an effective negative braking index,  possibly related to permanent $\dot \nu$ offsets associated with the  glitches.} of J0537-6910. We do not have to bend our understanding of the physics very much to make the observations fit the theory. Given this,  let us consider the problem from an observational point-of-view.
To be specific; can we use observations to constrain our  ignorance about the theory?

There are two (obvious) ways to address this question. Additional X-ray timing of the pulsar may strengthen (or not) the argument in favour of a braking index close to $n=7$. This would be further evidence in favour of the r-mode scenario, but it would still be circumstantial. Meanwhile, a dedicated gravitational-wave search may provide a limit on the allowed r-mode amplitude. Since we require the gravitational-wave emission to dictate the observed spin down one might be able to set a strong enough constraint to rule this out. In reality, the two kinds of observations are linked. In order to achieve the best gravitational-wave sensitivity one would need a reliable timing solution, e.g. provided by NICER. In absence of this one would have to fall back on a less optimal search strategy. 

In order to set the stage for a more detailed discussion of the detection problem, we assess the detectability of the emerging gravitational waves in the standard way.
First of all, we note that  (ignoring relativistic correction, see below)
the frequency of the emerging gravitational waves is
(for the main $l=m=2$ r-mode)
\begin{equation}
f_{\rm gw} = {4\nu \over 3} \approx 83~\mathrm{Hz} \ .
\end{equation}
We combine this with the gravitational-wave flux formula, making use of the  idealised source-detector configuration used for deformed spinning stars \citep{anna,ben}. That is, we use (assuming an optimal orientation)
\beq
h_0^2 = {10G \over c^3} \left({ 1\over 2\pi f_\mathrm{gw} d}\right)^2 \dot E
\label{rflux}
\eeq
where $d$ is the distance to the source. Combining this with the gravitational-wave luminosity for the r-modes, we arrive at
\beq
h_0 \approx {3\alpha_s \over 4 d}   \left( {10 GMR^2 \tilde J\over c^3 t_\mathrm{gw}}\right)^{1/2}
\eeq
Scaling to suitable parameter values, we have
\beq
h_0 \approx 7.5\times 10^{-25} \alpha_s \left( {M\over 1.4M_\odot}\right) \left( {R\over 10\ \mathrm{km}}\right)^3 \left( {\nu \over 100\ \mathrm{Hz}}\right)^3 \left( {50\ \mathrm{kpc} \over d} \right)
\label{hoft}
\eeq
Assuming that the r-mode amplitude is, indeed, the $\alpha_s$ inferred from the spin down and that the distance to the pulsar is 50~kpc, we have $h_0 \approx 2-3 \times 10^{-26}$ for a neutron star radius in the range $10-14$~km.

As a rough idea of the detectability of this signal, let us assume that the system does not evolve much during the observation period (the frequent glitches may be a problem). Then the effective amplitude increases as the square root of the observing time $t_\mathrm{obs}$ and we simply assess the detectability by comparing $\sqrt{t_\mathrm{obs}} h_0(t)$  to $11.4\sqrt{S_n}$, where $S_n$ is the power spectrum of the detector noise.

For a targeted search, the comparison we need is, in fact, straightforward. We can use the most recent targeted search for continuous gravitational waves, based on about 70 days of LIGO data from the first observing run (O1), from \citet{O1target}. Estimating the sensitivity at $f_\mathrm{gw}\approx 80$~Hz from Figure~1 in that paper, we see that $h_0 \approx 3\times 10^{-26}$, very close to our estimated strain. This is obviously interesting. Of course, we need to do a little bit better to rule out (or in!) the scenario. If we instead consider advanced LIGO operating at design sensitivity, then we would have
$\sqrt{S_n} \approx 4\times10^{-24}$ at the frequency we are interested in (see figures in \citet{livrev}). From this we see that the predicted level of signal would be detectable with less than 2 months worth of data. With a longer integration time one might be able to put interesting  constraints on the model.
 
Of course, these estimates assume a targeted search, which in turn requires a reliable timing solution. Given that it is not clear that this information will be available, it is worth considering how well one may be able to do with a less optimal strategy. Since we know the location of the source, the natural strategy would be a directed search similar to that used for  supernova remnants \citep{casa}. In this case, the attainable sensitivity is not at the level we have assumed. We get an idea of what is achievable by comparing the LIGO results from the 6th Science run (S6), e.g. the results from \citet{casa} and \citet{s6allsky} (for reference, the best current all sky search results, based on O1 data, can be found in \citet{O1allsky} and \citet{O1allsky2}). Very roughly, this comparison suggests that one would lose about one order of magnitude of sensitivity in a blind all-sky search, but gain a factor of two or so back in a directed search. Assuming that one would lose a factor of 5 in sensitivity, one would have to compensate by increasing the effective integration time by a factor of 25. This likely makes the required observation time prohibitively long even for advanced LIGO at design sensitivity. Such a search may require a future generation of instruments. 

Nevertheless, we should (eventually) be able to use observations to either confirm or rule out (which may be more likely) the notion that an unstable r-mode is present in J0537-6910. Of course, in order to carry out the suggested gravitational-wave search one would need to go a couple of steps beyond our simple estimates. Perhaps most importantly, one has to consider the fact that the true r-mode frequency is not going to be $4\nu/3$. In a realistic neutron star model, the r-mode frequency is shifted by a range of effects. For a relatively slowly spinning star (such that one can ignore rotational shape corrections) the largest correction is likely due to relativity (the gravitational redshift and the rotational frame dragging \citep{lock1,lock2}, see also \citet{andho}). The most detailed investigation of the problem was presented in \citet{rmgr}. The results suggest that we  should consider the realistic r-mode frequency to lie in the range $1.39 \nu < f_\mathrm{gw} < 1.57 \nu$. That is, for J0537-6910 one should search for a signal in the range $f_\mathrm{gw}\approx 86-98$~Hz (note that the Newtonian result, $f_\mathrm{gw}\approx 83$~Hz, is not inside this interval).

It is also worth commenting on the challenge of carrying out a search for gravitational waves from J0537-6910 without timing data. Any such effort could be seriously affected by the frequent glitches \citep{ashton}. The glitches would also impact on attempts to stack shorter segments of data to increase the sensitivity. While one may, in principle, be able to stack inter-glitch data (shorter than the 3 month or so interval between glitches) this may be practically difficult without an identification of the glitches in the first place. In reality, it may be tricky to carry out the required search in archival O1-O2 LIGO data. 

\section{Concluding remarks}

The 62~Hz X-ray pulsar PSR J0537-6910 is undoubtedly an intriguing object. The fastest known young pulsar, it has a complex spin-history frequently interrupted by glitches. This makes matching timing observations to theory a real challenge. At the same time, one may hope that the enigmatic behaviour may shed light on the involved physics, like the superfluid vortex dynamics thought to dictate the relaxation after each glitch event. 

We have argued that  J0537-6910 should be a prime target for a joint observing campaign between NICER and the LIGO-Virgo network. The argument draws on an analysis of the complete timing data from RXTE and an observation that a trend in the inter-glitch behaviour of the pulsar may
indicate an effective braking index close to $n=7$. This value would accord with a neutron star spinning down due to gravitational waves from an unstable r-mode. We have discussed to what extent this scenario may be consistent and whether the associated gravitational-wave signal would be within reach of ground-based detectors. In essence, our estimates suggest that one may well be able to use observations to constrain (or even rule out) the idea. This is an interesting prospect for the future.

\section*{Acknowledgements}
We would like to thank Lucien Kuiper for deriving and providing the times-of-arrival used in this study. We are also grateful to Maria Alessandra Papa for helpful discussions. NA and WCGH acknowledge funding from STFC in the UK through grant number ST/M000931/1. DA acknowledges support from the Polish National Science Centre (SONATA BIS 2015/18/E/ST9/00577, P.I.: B. Haskell). CME acknowledges funding from CONICYT FONDECYT/Regular 1171421 in Chile. BH has received funding under the European Union's Horizon 2020 research and innovation programme under grant agreement No. 702713. Partial support comes from NewCompStar, COST Action MP1304.

\end{document}